\documentclass[prd,aps,twocolumn,a4paper,showkeys,nofootinbib,showpacs]{revtex4-1}

\usepackage{graphicx,psfrag}
\usepackage{mathrsfs}
\usepackage{amsmath,amsfonts,amssymb}
\usepackage{multirow}
\usepackage{comment}

\usepackage{xspace}
\usepackage{ulem}
\usepackage{hyperref}
\usepackage{enumitem}

\newcommand{\be}{\begin{equation}}
\newcommand{\ee}{\end{equation}}
\newcommand{\bea}{\begin{eqnarray}}
\newcommand{\eea}{\end{eqnarray}}
\newcommand{\bel}{\begin{align}}
\newcommand{\eel}{\end{align}}

\newcommand{\tGRAthena}{\texttt{GR-Athena++}}
\newcommand{\GRAthena}{\tGRAthena\xspace}

\newcommand{\tAthena}{\texttt{Athena++}}
\newcommand{\Athena}{\tAthena\xspace}

\newcommand{\tTwoPunctures}{\texttt{TwoPunctures}}
\newcommand{\TwoPunctures}{\tTwoPunctures\xspace}

\def\Msun{{\rm M_{\odot}}}

\def\GMc2{{\rm G M_{\odot} c^{-2}}}

\def\l{\ell}

\def\kt2{\kappa^\text{T}_2}

\def\Mo{{\rm M_{\odot}}}
\def\kt2{\kappa^\text{T}_2}

\def\Ng{\mathcal{N}_{\mathrm{g}}}
\def\Ncg{\mathcal{N}_{\mathrm{cg}}}

\def\2nd{2^\mathrm{nd}}
\def\4th{4^\mathrm{th}}
\def\6th{6^\mathrm{th}}
\def\8th{8^\mathrm{th}}
\def\sp{\delta x}

\def\z4c{$\mathrm{Z}4\mathrm{c}$}
\def\z4oc{$\mathrm{Z}4(\mathrm{c})$}
\def\z4{$\mathrm{Z}4$}
\def\ccz4{$\mathrm{CCZ}4$}

\newcommand{\Mesh}{\texttt{Mesh}}
\newcommand{\MeshBlock}{\texttt{MeshBlock}}

\usepackage{color}
\definecolor{cyan}{rgb}{0,0.9,0.9}
\definecolor{orange}{rgb}{0.9,0.5,0}
\definecolor{purple}{rgb}{0.8,0.4,0.8}
\definecolor{grey}{rgb}{0.8242,0.8242,0.8242}
\definecolor{brickred}{rgb}{0.8, 0.25, 0.33}
\definecolor{magenta}{rgb}{1,0,1}

\begin{document}

\title{Numerical relativity simulations of compact binaries:
  comparison of cell- and vertex-centered adaptive meshes}

\author{Boris \surname{Daszuta}$^{1}$}
\author{William \surname{Cook}$^{1}$}
\author{Peter \surname{Hammond}$^{2}$}
\author{Jacob \surname{Fields}$^{2,3}$}
\author{Eduardo M. \surname{Guti\'errez}$^{2,3}$}
\author{Sebastiano \surname{Bernuzzi}$^{1}$}
\author{David \surname{Radice}$^{2,3,4}$}
\affiliation{${}^1$Theoretisch-Physikalisches Institut, Friedrich-Schiller-Universit{\"a}t Jena, 07743, Jena, Germany}
\affiliation{${}^2$Institute for Gravitation and the Cosmos, The Pennsylvania State University, University Park, PA 16802, USA}
\affiliation{${}^3$Department of Physics, The Pennsylvania State University, University Park, PA 16802, USA}
\affiliation{${}^4$Department of Astronomy \& Astrophysics, The Pennsylvania State University, University Park, PA 16802, USA}

\date{\today}

\begin{abstract}
  Given the compact binary evolution problem of numerical relativity, in the finite-difference, block-based, adaptive mesh refinement context, choices must be made on how evolved fields are to be discretized. In \tGRAthena{}, the space-time solver was previously fixed to be vertex-centered. Here, our recent extensions to a cell-centered treatment, are described. Simplifications in the handling of variables during the treatment of general relativistic magneto-hydrodynamical (GRMHD) evolution are found. A novelty is that performance comparison for the two choices of grid sampling is made within a single code-base. In the case of a binary black hole inspiral-merger problem, by evolving geometric fields on vertex-centers, an average $\sim 20\%$ speed increase is observed, when compared against cell-centered sampling. The opposite occurs in the GRMHD setting. A binary neutron star inspiral-merger-collapse problem, representative of typical production simulations is considered. We find that cell-centered sampling for the space-time solver improves performance, by a similar factor. 
\end{abstract}


\maketitle

\section{Introduction}
\label{sec:intro}

Fashioning accurate description of astrophysically sourced, binary constituent encounters, under extreme conditions, is a pressing challenge for the numerical relativity (NR) community. Multimessenger detection efforts \cite{TheLIGOScientific:2017qsa,Goldstein:2017mmi,Savchenko:2017ffs} benefit when supplemented by such data. Observational efforts of next-generation gravitational-wave detectors \cite{akutsu2020overviewkagracalibration,amaroseoane2017laserinterferometerspace,Punturo:2010zz,Evans:2016mbw} will also greatly benefit from higher fidelity reference numerical data, that spans a larger range of parameter space (or e.g.~models \cite{Nagar:2018zoe} suitably informed by). To confront simulation challenges, sophisticated general relativistic, magneto-hydrodynamical (GRMHD) techniques (see e.g.~\cite{Font:2007zz}), and microphysics descriptions \cite{Shibata:2011kx,Radice:2021jtw} have been pursued. In tandem, the development of code infrastructure, and algorithms, with the aim of improved performance scaling on modern high-performance computing (HPC) resources, has also received recent, particular attention. Some examples of modern NR codes under active development with these goals include: %
{\tt bamps} \cite{Hilditch:2015aba,Bugner:2015gqa}, %
{\tt Dendro-GR} \cite{Fernando:2018mov}, %
{\tt Einstein Toolkit} \cite{Loffler:2011ay} ({\tt GRaM-X} \cite{Shankar:2022ful}), %
{\tt GRChombo} \cite{Clough:2015sqa}, %
{\tt NMesh} \cite{tichy2023newdiscontinuousgalerkin}, %
{\tt SpECTRE} \cite{Kidder:2016hev}, %
and {\tt SPHINCS\_BSSN} \cite{Rosswog:2020kwm}. %
A variety of numerical methods \cite{Grandclement:2007sb,Doulis:2022vkx,Alfieri:2018a}, algorithmic approaches \cite{stout1997adaptiveblockshigh,Berger:1984zza,burstedde2019numberfaceconnectedcomponents,morton1966computer}, and NR formulations %
\cite{Nakamura:1987zz,Shibata:1995we,Baumgarte:1998te,%
Bernuzzi:2009ex,Hilditch:2012fp,%
Friedrich:1985,Pretorius:2005gq,Lindblom:2005qh} %
are employed, to list but a few prominent examples. In spite of this, the actual time taken (HPC resource consumed) in the calculation of numerical evolutions, particularly for extreme scenarios that require highly resolved physical scales, can require weeks, or months \cite{Lousto:2020tnb}. Clearly, even minor performance optimizations, of a few percent, can have a significant impact on such HPC wait times, together with broader, sustainability considerations \cite{lannelongue2023greenerprinciplesenvironmentally}.

In this work, we focus on finite-difference based evolution of the \z4c{} formulation \cite{Bernuzzi:2009ex,Hilditch:2012fp} of NR, as coupled to the general relativistic, hydrodynamical (GRHD) formulation\footnote{For simplicity we do not investigate magnetic fields in this work.} of \cite{Banyuls:1997zz}. This is precisely the setting treated by our code \GRAthena{} \cite{Daszuta:2021ecf,Cook:2023bag} (see also \cite{daszuta2024grathenamagnetohydrodynamicalevolution}). In brief, this is our extension to the astrophysical (radiation), GRMHD code \Athena{} \cite{white2016extensionathenacode,%
felker2018fourthorderaccuratefinite,%
stone2020athenamathplusmathplus}, we have geared toward evolution with dynamical space-time. Building upon the underlying octree, block-based adaptive-mesh-refinement (AMR) infrastructure, with addition of our \z4c{} based space-time solver both in vacuum \cite{Daszuta:2021ecf}, and for GRMHD with dynamical space-time \cite{Cook:2023bag}, has led to demonstrable scaling efficiencies in excess of $80\%$ on (pre-)exascale HPC infrastructure.

Specifically, here, we aim to address a seemingly innocuous question: should we select the geometric fields, that is, the dynamical fields of \z4c{}, to be discretized on vertex-centers (VC) or cell-centers? Our initial choice was based on considerations for binary black hole evolution, where transfer of data between differing levels of refinement, is less computationally expensive for VC in contrast to cell-centers. On the other hand, in consideration of GR(M)HD, typically hydrodynamical sampling is on cell-centers (see e.g.~{\tt BAM} \cite{Brugmann:2008zz}, {\tt GRHydro} \cite{Moesta:2013dna}, and {\tt WhiskyTHC} \cite{Radice:2013xpa}). This choice is in part due to greater amenability of the underlying hydrodynamical treatment as based on conservative, high-resolution-shock-capture (HRSC) methods (see e.g.~\cite{Thierfelder:2011yi}). \Athena{} also handles hydrodynamical description on cell-centers. Having previously fixed geometric fields to VC, discretized fields must consequently be transferred between grids, which incurs an overhead. A priori it is not necessarily clear as to how large this is. Our aim here is to investigate switching to a cell-centered treatment of the space-time solver in \GRAthena{}, thus providing an answer. Such a code-internal, investigation, does not appear to have been previously performed. This gap we fill here. 
This paper proceeds as follows. In \S\ref{sec:method} we briefly explain the computational domain structure and the variable discretization strategy. Subsequently, we present a simplified refinement strategy, tailored to the binary merger problem, and independent of geometric field sampling selection. In \S\ref{sec:val} we demonstrate the robustness of the technique through numerically solving BBH and BNS inspiral-merger problems, representative of typical production simulations. Section \S\ref{sec:discussion} concludes.

\section{Method}
\label{sec:method}

\GRAthena{} \cite{Daszuta:2021ecf,Cook:2023bag} builds upon \Athena{} \cite{white2016extensionathenacode,%
felker2018fourthorderaccuratefinite,%
stone2020athenamathplusmathplus} and thus inherits overall features of the latter framework. In brief, problems are formulated over a target computational domain $\Omega$ (the \Mesh{}) which is partitioned into a collection of sub-domains $\Omega_i$ (\MeshBlock{} objects). We have $\Omega = \sqcup_{i\in Z} \Omega_i$ where $Z$ may be thought of as an ordered parametrization of multi-dimensional coordinates to a space-filling curve \cite{morton1966computer,burstedde2019numberfaceconnectedcomponents}. For a $d$-dimensional problem the overall extent of $\Omega$ together with suitable boundary data on $\partial \Omega$ must be provided. Following this, the underlying discretization is fixed through specification of \Mesh{} and \MeshBlock{} sampling $N_M=(N_{M_1},\,\dots,\,N_{M_d})$ and $N_B=(N_{B_1},\,\dots,\,N_{B_d})$, respectively, where $N_B$ must divide $N_M$ component-wise\footnote{For a uniform number of samples in each dimension we typically use a single scalar to represent all component values of the tuples.}. In the case that (local) refinement is desired, \MeshBlock{} partitioning into $2^d$ new \MeshBlock{} objects is recursively performed. This involves halving size along each axis at a fixed number of samples, which doubles the resolution at each step. The aforementioned is done while maintaining an (at-most) $2:1$ refinement ratio between neighboring $\{\Omega_i\}_{i\in Z}$. While the logical-rectilinear structure here described may then be further mapped to suitable curvilinear coordinates, for simplicity we focus on the Cartesian context.
\subsection{Cell-centered extension}
\label{sbsec:cx_meth}

In \GRAthena \cite{Daszuta:2021ecf,Cook:2023bag} we extended sampling support to enable vertex-centered (VC) discretization of field variables together with infrastructure enabling high-order approximations to be carried out. In this work we focus on analogous extensions in description of geometric quantities on cell-centers (CX). A variable $\mathcal{V}(x)$ where $x\in[a,\,b]$ $(a<b)$ is CX discretized\footnote{The grid sampling will also be referred to as CC. On CX fields are to be thought of as pointwise sampled, whereas on CC they carry a cell-averaged interpretation.} to $\mathcal{V}_{i+1/2}$ over $x_{i+1/2}:=a+(i+1/2)\sp$ where $\sp{}:=(b-a)/N_B$ and $i=0,\,\ldots,\,N_B-1$. This yields $N_B$ total samples. In practice the interval is extended by an additional $\Ng{}$ ``ghost'' nodes  either-side\footnote{In the case that $d=\dim(\Omega)>1$ we utilize a tensor product grid of such extended one-dimensional discretizations.}. This extension facilitates evaluation of e.g.~derivative approximants, transfer of field data between neighboring \MeshBlock{} objects that may differ in the level of refinement, and/or imposition of boundary conditions.

\begin{figure}[!ht]
	\centering
		\includegraphics[width=\columnwidth]{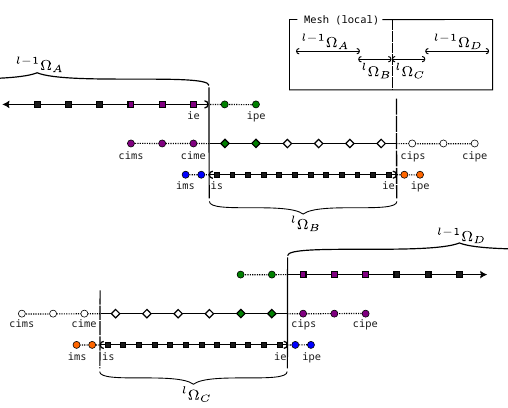}
    \caption{%
      One-dimensional illustration of inter-\MeshBlock{} communication given a refined local \Mesh{} structure. Each \MeshBlock{} features a fundamental representation of a sampled field depicted through square nodes which is extended in each direction through ghost nodes (circles) that facilitate communication. Additionally a complementary coarse sampling is present which is explicitly depicted for data on level $l$ and suppressed on level $l-1$. Here $N_B=12$, $\Ng{}=2$, and $\Ncg{}=3$. Prior to all communication data in filled square nodes is available; the task is thus to fill data at locations indicated by shaded circles and diamonds. Nodes are assigned \MeshBlock{}-local labels for ease of discussion.
    }
		\label{fig:level_to_level}
\end{figure}

In order to populate ghost nodes, communication is required. In the case that $\mathcal{V}$ is discretized over a \Mesh{} decomposed at a fixed,  common level, under CX sampling, this entails that nearest-neighbor \MeshBlock{} objects can be utilized to populate ghost-layers through a straight-forward copy. Suppose instead that sub-domains differ in level $l$. In this situation, analogous to the VC case, each sub-domain ${}^l\Omega_i$ (where prefix $l$ indicates level) as described by a \MeshBlock{}, features a fundamental and an additional coarse representation of the sampled variable. This latter is one level above at $l-1$. Thus if $N_B$ is the number of physical points on a \MeshBlock{} extended by $2\Ng{}$ ghost nodes, then the complementary coarse representation features $N_B/2$ physical points extended by $2\Ncg{}$ coarse ghosts. The situation is illustrated (locally) in Fig.\ref{fig:level_to_level}.

The general communication strategy and ordering of operations closely follows and extends that of \Athena{}. This minimizes the amount of infrastructure changes required. Synchronization of data across levels may be understood and implemented by considering the transfer operations: fine to coarse (restriction $\mathcal{R}$), and coarse to fine (prolongation $\mathcal{P}$). As an example consider synchronization of ghost-layer data between ${}^{l-1}\Omega_A$ and ${}^{l}\Omega_B$ as depicted in Fig.\ref{fig:level_to_level}. \textit{Initially} data on the fundamental grid of ${}^{l}\Omega_B$ over the nodes $(\texttt{is},\,\texttt{is}+1,\,\ldots,\,\texttt{ie})$ is known. This is first restricted to a subset of the complementary coarse grid of ${}^{l}\Omega_B$ over the nodes $(\texttt{cime}+1,\,\ldots,\,\texttt{cime}+\Ng{})$; see green diamonds in the figure. Subsequently data at the nodes $(\texttt{ie}+1,\,\ldots,\,\texttt{ipe})$ of ${}^{l-1}\Omega_A$ may be filled through communication or a copy (green circles in figure). Conversely using data from ${}^{l-1}\Omega_A$ at the nodes $(\texttt{ie}-\Ncg{}+1,\,\ldots,\,\texttt{ie})$ allows for filling of $(\texttt{cims},\,\ldots,\,\texttt{cime})$ on ${}^{l}\Omega_B$ (purple circles) through communication or a copy. Finally, ghost layer nodes $(\texttt{ims},\,\ldots,\,\texttt{is}-1)$ of the fundamental grid over ${}^{l}\Omega_B$ may be filled (blue circles) through $\mathcal{P}$ based on data local to the \MeshBlock{}. The procedure is analogous for the pair ${}^{l}\Omega_C$ and ${}^{l-1}\Omega_D$.

Comparing node positions in Fig.\ref{fig:level_to_level} across differing levels of refinement clearly shows that nodes do not align but are fully staggered\footnote{One could instead consider $3:1$ refinement ratios to obviate this.}. This is one aspect where CX differs from VC sampling \cite{Daszuta:2021ecf}. The coarsening and refining of sampled field data via $\mathcal{R}$ and $\mathcal{P}$ respectively, is based on polynomial interpolation under CX and VC. Due to the inter-level node staggering occuring in CX, a non-trivial evaluation of interpolants for every target node is required. This is in contrast to the case of VC where a subset of operations reduces to injection (copy) \cite{Daszuta:2021ecf}. The distinction heralds a performance implication as will be demonstrated in \S\ref{sbsec:vac_tests} and \S\ref{sbsec:mat_tests}.

We close this subsection with a collection of technical points concerning implementation aspects.

\begin{itemize}
  \item{The formal order of accuracy for the $\mathcal{R}$ and $\mathcal{P}$ operations on CX is controlled by the number of nodes selected. We fix this choice based on $\mathcal{N}_g$ selection so to match the approximation order with that of the finite-differencing.}
  \item{Given a target node, source data entering an interpolant describing CX $\mathcal{P}$ has a directional bias due to grid staggering. This can be understood by noting that the source data nodes are not symmetrically placed about the location of the target node, but offset. Consider at a desired formal order of accuracy, the initial coarse, complementary restriction needed when communicating data to a finer neighboring \MeshBlock{}. There may be insufficient nodes available to evaluate with symmetrically spaced source data. In this situation, we again bias the stencil selection.   }
  \item{As is well-known, floating point (FP) arithmetic is not necessarily associative: $(a+b)+c\neq a+(b+c)$ for FP numbers $\{a,\,b,\,c\}$. If care is not taken in how numerical expressions are ordered, errors can e.g.~lead to catastrophic cancellations, or result in spurious symmetry breaking during simulations \cite{fleischmann2019numericalsymmetrypreservingtechniques}. A combination of Kahan-Babuska-Neumaier compensated summation \cite{neumaier1974rundungsfehleranalyseeinigerverfahren}, and explicit symmetrization of arithmetic operations involving averaging can partially mitigate this \cite{fleischmann2019numericalsymmetrypreservingtechniques}. This can be optionally activated for interpolant evaluation. We have not however found it necessary for the simulations presented here.}
  \item{There are now two styles of cell-centered sampling where discretized field variables may be registered and stored: that inherited from \Athena{}, which we denote CC, and the newly introduced CX. The principal distinction is that for a \Mesh{} featuring \MeshBlock{} objects on distinct levels, the $\mathcal{R}$ and $\mathcal{P}$ operators are distinct: for CC there is minmod based slope-limiting \cite{Stone:2020}, whereas for CX we utilize the Barycentric Lagrange approach \cite{berrut2004barycentriclagrangeinterpolation}. Special care has been taken to ensure that distinct fields may be suitably registered as either CC or CX with their respective, logic available concurrently.}
  \item{The VC double-restriction operation described in \cite{Daszuta:2021ecf} we have found to not be strictly required. Its removal means that when utilizing vertex-centered sampling we now have $N_B\geq \max(4,\,\mathcal{N}_g)$. This allows for smaller \MeshBlock{} size, which offers greater flexibility in tailoring resolution adapted regions.}
\end{itemize}

\subsection{Refinement strategy}
\label{sbsec:meth_ref}

In order to resolve dynamical features over widely disparate spatial scales we utilize adaptive mesh refinement (AMR). Our purpose here is to fix notation, and to point to two simple, minimal criteria that will facilitate consistency tests of the new infrastructure. The aforementioned feature evolution of initial data describing a binary black hole (BBH) (\S\ref{sbsec:vac_tests}), and a binary neutron star  (\S\ref{sbsec:mat_tests}). For a comprehensive investigation of AMR criteria in \GRAthena{}, and effect on waveform quality for BBH evolution, see \cite{Rashti:2023wfe}.

The AMR conditions of \GRAthena{} are quite flexible. A user-specified function is defined for each MeshBlock; this function returns a flag that controls whether to (de)-refine. Consider puncture-based evolution \cite{Brugmann:2008zz} of a BBH system. Here a black hole center is described through a puncture. The position of a puncture $\mathbf{x}_p(t)$ can be computed by augmenting the dynamical system evolved through an additional ODE based on gauge information \cite{Campanelli:2005dd}. In short, this is one variety of a so-called tracker. This allows us to construct a dynamically adjusted, octree-based, pseudo-box-in-box (pBIB) refined \Mesh{} as detailed in \cite{Daszuta:2021ecf} (or $L_\infty$-norm refinement in the language of \cite{Rashti:2023wfe}).

Let us suppose the \Mesh{} is the Cartesian coordinatized, computational domain $\Omega:=[-D,\,D]^3$. A parameter controlling the maximal level of refinement $N_L$ is specified. A hierarchy of \MeshBlock{} objects is then constructed, which increases in resolution as a puncture is approached (see e.g.~Fig.\ref{fig:mesh_calib_110}). On the finest resolution level the grid spacing will be $\delta x=2D / (N_M 2^{N_L-1})$ \cite{Daszuta:2021ecf}. As an alternative to pBIB, we can exploit information from the trackers and directly impose a desired region to be at fixed resolution (equivalently level). Denote a spherical region of radius $R$, centered at $p$ as $\mathbb{S}_R(p)$. We center such a region on a puncture and impose a desired local level of refinement $l$. This we write as $\mathbb{S}_{l,R}(\mathbf{x}_p(t))$.

In a similar vein, consider a dynamical, user selected, control field $\chi$ featuring a local extrema $\chi^*$. Another type of tracker is possible to construct based on following the time development of $\chi^*(t)$ through $\mathbb{S}_{l,R}(\chi^*)$. This allows for the region surrounding the extrema to be captured as its position changes in time. If multiple such regions are registered, occur at differing $l$, and happen to intersect, then we take the finest level as the target level. This is the principal condition that we employ for AMR during BNS evolution.

\section{Validation}
\label{sec:val}
Our first goal in this section is validating the extended cell-centered implementation in the context of vacuum binary black hole evolution. The space-time solver implemented in \GRAthena{} is based on the \z4c{} formulation \cite{Bernuzzi:2009ex,Hilditch:2012fp}. Here the space-time is constructed based on the evolution of the dynamical, geometric variables $\mathcal{Z}:=\big(%
\chi,\,\tilde{\gamma}{}_{ij},\,\hat{K},\tilde{A}{}_{ij},\,\Theta,\,\tilde{\Gamma}{}^i%
\big)$. The definition and meaning of these variables is standard and may be found, together with implementation details, in \cite{Daszuta:2021ecf}. Previously, motivated by AMR efficiency considerations in the vacuum context, we have chosen these to evolve over VC grid sampling. We compactly denote this through $\mathcal{Z}_{\mathrm{VC}}$. This work details a relaxation of this restriction, extending functionality to allow for CX sampling, i.e.~sampling and evolving directly as $\mathcal{Z}_{\mathrm{CX}}$. We detail tests of this in \S\ref{sbsec:vac_tests}.

Secondly, we investigate the binary neutron star (BNS) inspiral-merger problem. Our recently implemented GRMHD treatment in \GRAthena{} is based on the formulation of \cite{Banyuls:1997zz}. For full details, and conventions, see \cite{Cook:2023bag}. Here we recall key ingredients of constructing the matter side of the evolution as based on selecting $\mathcal{Z}_{\mathrm{VC}}$. For simplicity, we restrict our exposition to the GRHD sector. For the hydrodynamics, the (conserved) field variables are $\mathcal{Q}_A:=\big(D,\,S{}_j,\,\tau\big)$, %
and satisfy a balance law. Recall that this is evolved based on conservative, finite-volume, HRSC techniques \cite{Thierfelder:2011yi}. The natural choice of sampling, that also leverages extant \Athena{} infrastructure is cell-centered, i.e. we take~$\mathcal{Q}_{A,\mathrm{CC}}$. We also work with a set of complementary, primitive variables $\mathcal{R}_A:=\big(\rho,\,\tilde{u}{}^i,\, p\big)$. The relation $\mathcal{Q}_A(\mathcal{R})$ is non-linearly implicit, and is evaluated through numerical inversion \cite{Noble:2005gf}. In this work the conservative to primitive variable mapping is handled through the use of {\tt RePrimAnd} \cite{Kastaun:2020uxr}. This inversion requires geometric fields to be available on the same grid sampling as the matter. Hence an intergrid interpolation $\mathcal{Z}_{\mathrm{VC}}\rightarrow \mathcal{Z}_{\mathrm{CC}}$ is required.

In order to evolve the quantities $\mathcal{Q}_{A,\mathrm{CC}}$ via HRSC we require assembly of fluxes at cell-interfaces (that is, face-centers (FC)). In this work, this is achieved based on the reconstruction of $\mathcal{R}_{A,\mathrm{CC}}$ to FC with the fifth order method of \cite{Borges:2008a} denoted WENO5Z. Flux assembly also requires a subset of $\mathcal{Z}_{\mathrm{FC}}$ and hence another interpolation. The interface fluxes are combined according to the Local-Lax-Friedrichs (LLF) scheme (see e.g.~\cite{zanna2002efficientshockcapturingcentraltype}). Finally the sources appearing in the hydrodynamical balance law are assembled, which again requires $\mathcal{Z}_{\mathrm{CC}}$. On the other hand, the gravitational matter density, momenta, and stress quantities that recouple the system to the space-time solver, must be provided on VC. This entails the interpolation of hydrodynamical quantities to VC. It is clear from the above summary, that by instead taking the evolved geometric variables as $\mathcal{Z}_{\mathrm{CX}}$ from the start, fewer transfers of geometric data between grids would be required. We further validate our code functionality, for BNS, in \S\ref{sbsec:mat_tests}. 

In addition to ensuring consistency of the implementation, and that problems are solved correctly, a practical performance assessment is needed. We demonstrate that execution speed, and suitable choice of $\mathcal{Z}_{\mathrm{VC}}$ ($\mathcal{Z}_{\mathrm{CX}}$) is contingent on the class of physical problem simulated.

\subsection{Binary Black Holes}
\label{sbsec:vac_tests}

We assess the performance of our code by focusing on the well-known calibration binary black hole (BBH) problem of \cite{Brugmann:2008zz}. The \z4c{} system is coupled to the moving puncture gauge using the $1+\log$-slicing condition \cite{Bona:1994a} and the Gamma driver shift \cite{Alcubierre:2002kk,vanMeter:2006vi}, with parameters as described in \cite{Daszuta:2021ecf}. Initial data is generated through the use of the \TwoPunctures\footnote{See \url{https://bitbucket.org/bernuzzi/twopuncturesc/}.} library based on \cite{Ansorg:2004ds}. We model two initial non-spinning black holes of bare-mass $m_\pm = 0.483~M$, located on the $x-$axis, with $x^1_\mathrm{p,\pm}(t=0) = \pm 3.257~M$, and initial momenta directed along the $y-$axis, $p^2_\mathrm{\pm}(t=0) = \mp 0.133~M$. These parameter choices lead to an evolution featuring $\sim 3$ orbits to merger, which occurs at an evolution time $t\sim 170~M$.

In these tests the Courant-Friedrich-Lewy (CFL) is fixed at $\mathcal{C}_1=0.2$. This choice determines the global time-step taken on each \MeshBlock{} based on the finest grid-spacing (occuring in the vicinity of a puncture and denoted $\sp_p$). Kreiss-Oliger dissipation is selected as $\sigma_{\mathrm{D}}=0.02$ throughout. The \z4c{} constraint damping parameters are taken as $(\kappa_1,\,\kappa_2)=(0.02,\,0)$ (see \cite{Daszuta:2021ecf}). Unless otherwise stated, in this section we take $\mathcal{N}_{g}=\mathcal{N}_{cg}$, that is, the number of ghosts on the fundamental grid is set equal to the number on the complementary coarse representation (see Fig.\ref{fig:level_to_level}). We select $\mathcal{N}_{g}=4$, which induces a formal $\6th$ order of spatial approximation. Time-evolution is
performed using the explicit $\4th$ order RK$4()4[2S]$ \cite{ketcheson2010rungekuttamethods}.

\begin{figure}[!ht]
	\centering
		\includegraphics[width=\columnwidth]{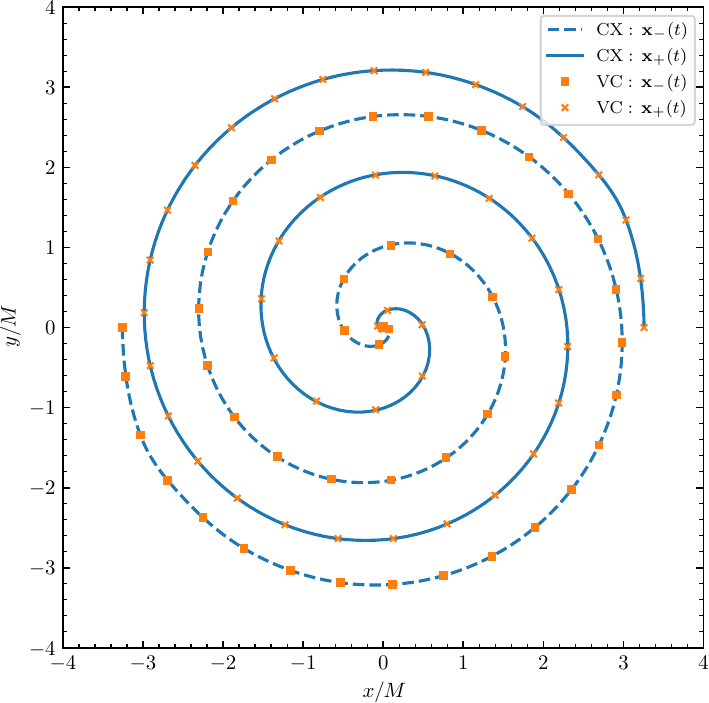}
    \caption{%
    Tracker evolution during calibration BBH inspiral and merger. The one-parameter families $\mathbf{x}_\pm(t)$ describe where punctures are located. Consistent behaviour for cell-centered extended (CX) and vertex-centered (VC) choices of sampling is observed.
    Parameters: $N_M=96$, and $N_B=16$. %
    }
		\label{fig:tracker_cx_vc}
\end{figure}

The overall grid configurations based on pBIB refinement (see \S\ref{sbsec:meth_ref}) are summarized in Tab.(\ref{tab:CalibBBHgridsLinf}).
\begin{table}[ht!]
  \centering
  \setlength\tabcolsep{2pt}
  \begin{tabular}{| c || c || c |}
    \hline
    $N_M$  & $\sp_p \times 10^{-2} [M]$ & $\#\mathrm{MB}$\\
    \hline
    $96$      & $3.125$   & $5144$ \\
    $128$     & $2.344$   & $6336$ \\
    $160$     & $1.875$   & $18640$ \\
    $192$     & $1.563$   & $22392$ \\
    \hline
  \end{tabular}
  \caption{%
  Properties of grid configurations utilized for vacuum tests. The total number of \MeshBlock{} objects ($\#\mathrm{MB}$) at the initial time is indicated. %
    In all cases the \Mesh{} is the Cartesian coordinatized domain $\Omega:=[-D,\,D]^3$, where $D=1536 M$ is selected. This entails that grid spacing is uniform along each spatial axis. In each case we select the maximal refinement through $N_L=11$. The \MeshBlock{} sampling is selected as $N_B=16$. Resultant puncture resolutions $\sp_p$ are also provided.
  }
  \label{tab:CalibBBHgridsLinf}
\end{table}
In principle the calibration BBH problem possesses a underlying bitant symmetry (i.e.~reflection across the $z=0$ plane to which the co-orbit and merger is confined). This allows for an approximate halving of the total computational resources required for a given problem. When imposing bitant symmetry, $\#\mathrm{MB}$ as listed in Tab.\ref{tab:CalibBBHgridsLinf} is roughly halved, and in order to keep the grid spacing fixed, we halve the \Mesh{} sampling parameter along the $z$-axis to $N_M/2$. 
As an initial consistency check, we select the geometric sampling as $\mathcal{Z}_{\mathrm{VC}}$ and the $N_M=96$ grid of Tab.\ref{tab:CalibBBHgridsLinf}. We perform a comparable run with $\mathcal{Z}_{\mathrm{CX}}$, for which we inspect the puncture tracker evolution in Fig.\ref{fig:tracker_cx_vc}. Excellent agreement can be seen.

\begin{figure}[!ht]
	\centering
		\includegraphics[width=\columnwidth]{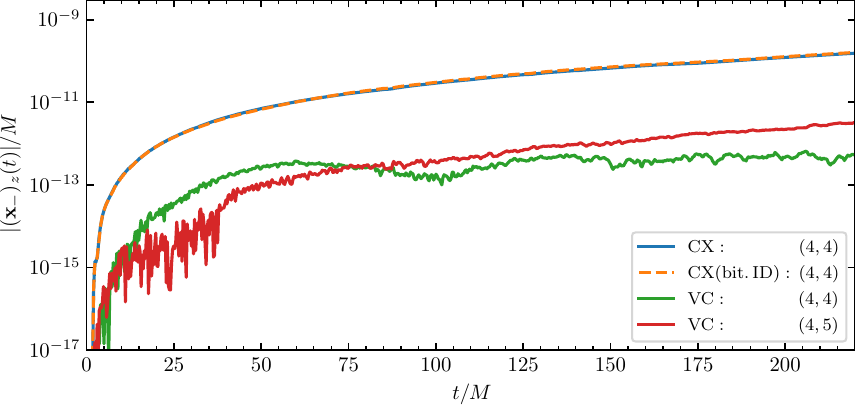}
    \caption{%
    Evolution of the $z$-component of the tracker $\mathbf{x}_-(t)$ during calibration BBH inspiral and merger. The CX and VC sampling choices are compared at fixed resolution. Parenthesis $(\mathcal{N}_{g},\,\mathcal{N}_{cg})$ indicate ghost node selection. The ``bit.ID'' set evolved with the choice $\mathcal{Z}_{\mathrm{CX}}$ has ID explicitly symmetrized. All runs of this figure are performed over full $3\mathrm{D}$ grids, regardless of ID symmetrization.
    Parameters: $N_M=96$, and $N_B=16$. %
    See text for further discussion.
    }
		\label{fig:tracker_cx_vc_z}
\end{figure}

The BBH evolution considered here should result in an evolution with a strict symmetry about the $z=0$ plane. During numerical evolution, however, this condition is not necessarily preserved. The $z$ component of a puncture can potentially slowly drift out of the plane. The behaviour is demonstrated in Fig.\ref{fig:tracker_cx_vc_z} for full $3\mathrm{D}$ runs. 
In the case of $\mathcal{Z}_{\mathrm{CX}}$, based on the structure of the level-to-level transfer operators, we do not observe, nor expect, strict FP symmetry preservation. Instead, during the full $3\mathrm{D}$ run we find a very minor drift developing, where the $z$-component of the puncture reaches $\sim\mathcal{O}(10^{-10})$. We also verify in Fig.\ref{fig:tracker_cx_vc_z} that $\mathcal{N}_{cg}$ may indeed be independently selected, so as to increase the formal order of accuracy in the prolongation operation for $\mathcal{P}$. We have similarly verified this for $\mathcal{Z}_{\mathrm{CX}}$. As this does not appear to substantively change simulation quality for the short evolution here, we proceed with runs that fix $\mathcal{N}_g=\mathcal{N}_{cg}=4$.

One reason for this behaviour, is that the \TwoPunctures{} library does not allow for explicit enforcement of an underlying symmetry about $z=0$. Thus, when interpolating an initial data (ID) set to a Cartesian grid, the result violates the bitant property at the FP level. We have taken special care to allow for symmetrization of the prepared ID by rewriting the interpolation process. The purpose here is to provide a test on preservation of FP symmetry across various elements of the infrastructure, such as, e.g.~operator approximants (finite-differencing kernels, interpolation, etc.).

In the case of $\mathcal{Z}_{\mathrm{VC}}$, explicit bitant symmetrization of the ID does lead to an evolution with \textit{exact} confinement of the punctures to the $z=0$ plane. This is in contrast to runs in Fig.\ref{fig:tracker_cx_vc_z}.

Having passed a tracker-based consistency check, we now check gravitational waveforms (GW). To do this, we construct the Weyl scalar $\Psi_4$ \cite{Brugmann:2008zz}, and then interpolate onto geodesic spheres \cite{Wang:2011}. Subsequently, we project through numerical quadrature evaluation, onto spin-weighted spherical harmonics \cite{Goldberg:1966uu} ${}_s Y_{lm}$ of spin-weight $s=-2$. Our convention differs from the reference by a Condon-Shortley phase of $(-1)^m$. The above allows us to extract $\psi{}_{lm}$. The $(\ell,\, m)$ modes of the gravitational wave strain $h$ satisfy $\psi_{\ell m}=\ddot{h}_{\ell m}$. The strain is then given by the mode-sum:
\begin{equation}
  R\left(h_+ - i h_\times\right) = \sum_{\ell=2}^{\infty}\sum_{m=-\l}^\l h_{\ell m}(t)\; {}_{-2}Y_{\ell m}(\vartheta,\varphi)\,.
\end{equation}
Following the convention of the LIGO algorithms library \cite{lalsuite} we set:
\begin{equation}\label{eq:strain_complex}
R h{}_{\ell m} = A{}_{\ell m} \exp(-i\phi{}_{\ell m}),
\end{equation}
and the gravitational-wave instantaneous frequency is defined through:
\begin{equation}
\omega{}_{\ell m}= \frac{d}{dt}\phi{}_{\ell m}.
\end{equation}
Note that the peak amplitude of the $(2,\,2)$ mode of $h$ is conventionally taken as the time of merger.
We denote this $u_{\mathrm{merger}}$.

\begin{figure}[!ht]
  \centering
    \includegraphics[width=\columnwidth]{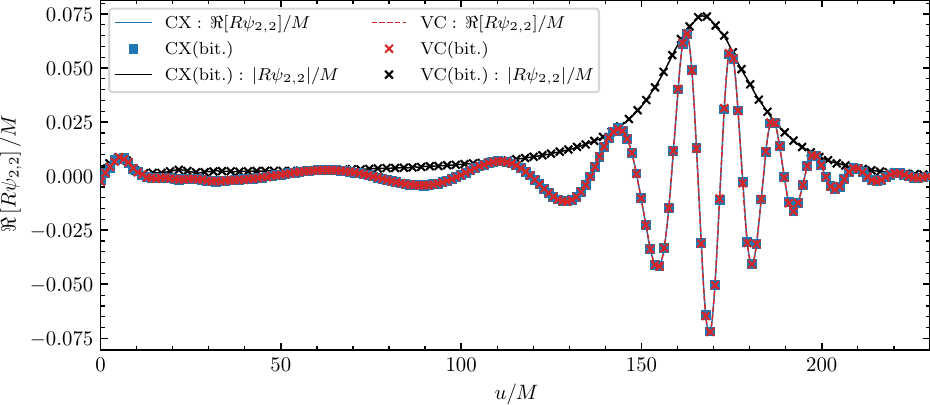}
    \caption{%
    Consistent behaviour between $\mathcal{Z}_{\mathrm{VC}}$ and $\mathcal{Z}_{\mathrm{CX}}$ for calibration BBH waveforms with and without bitant symmetry reduction. Black profiles depict the amplitude of the given mode based on calculations involving bitant symmetry reduction. Parameters: $N_M=96$, and $N_B=16$; extraction radius $R=85M$.     }
    \label{fig:wvf_cons_cx_vc}
\end{figure}

The quantity $\psi{}_{22}$ is illustrated for the Calibration BBH in Fig.\ref{fig:wvf_cons_cx_vc}.
Rather than just imposing ID bitant symmetrization, symmetry reduction on the $3\mathrm{D}$ grid is also imposed for a subset of runs. The waveforms continue to overlap consistently. This provides another important consistency test on implementation details.

\begin{figure}[!ht]
	\centering
		\includegraphics[width=\columnwidth]{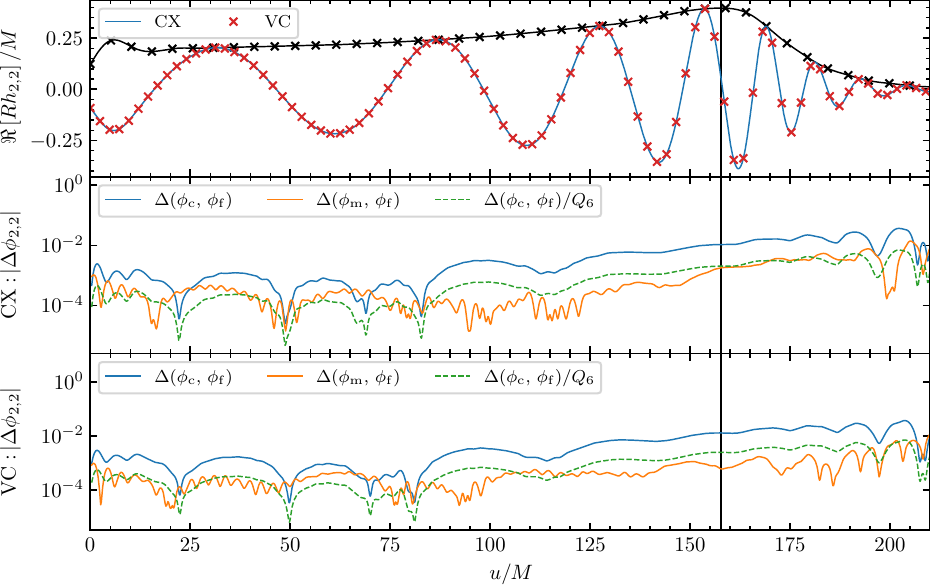}
    \caption{%
	    The $(2,\,2)$ mode of the gravitational strain (top-panel) together with convergence properties in the associated phase for CX and VC sampling in the middle and lower panels respectively. All quantities are taken as functions of the Schwarzschild tortoise coodinate $u$. We assess $\6th{}$ order compatible convergence. Indicated in a common vertical black line across panels is $u_{\mathrm{merger}}\simeq 158\,M$. We select calculations involving bitant symmetry reduction based on the coarse (c), medium (m), and fine (f) resolution triplet induced through selecting $N_M\in\{128,\,160,\,192\}$. Extraction performed at $R=85M$. 
    }
		\label{fig:wvf_conv_cx_vc_h_2_2}
\end{figure}

Beyond verification of qualitative behaviour, we next focus on the gravitational strain, impose bitant symmetry reduction, and assess convergence properties based on Appendix~\ref{sec:appendix}. The result, for constructing $h_{lm}$, based on time-domain integration (see e.g~\cite{albanesi2024scatteringdynamicalcapture}), is shown in Fig.\ref{fig:wvf_conv_cx_vc_h_2_2}.
A $\6th{}$ order compatible convergence trend is observed for the phase $\phi_{22}$. This is present for both $\mathcal{Z}_{\mathrm{VC}}$, and $\mathcal{Z}_{\mathrm{CX}}$. We have also verified that behaviour persists, consistently between grid sampling choices, for e.g.~the $(4,\,4)$ mode.

\begin{figure}[!ht]
	\centering
		\includegraphics[width=\columnwidth]{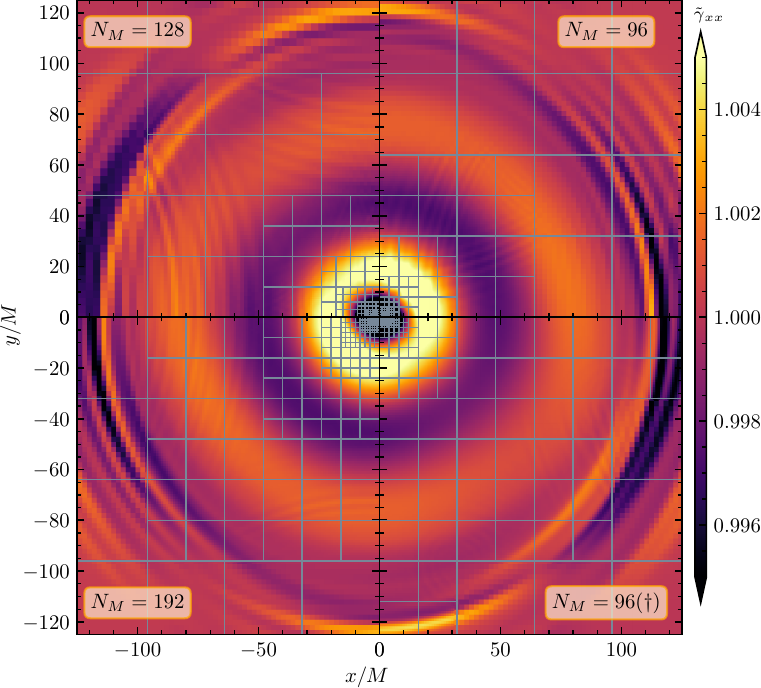}
    \caption{%
    \Mesh{} structure and values of $\tilde{\gamma}_{xx}$ under $\mathcal{Z}_{\mathrm{CX}}$ at $t\simeq110 M$ in the orbital plane. Each quadrant depicts distinct resolutions (AMR choices) for the same physical calibration problem configuration. Counter-clockwise, starting from the upper right quadrant we have pBIB AMR with $N_M\in\{96,\,128,\,192\}$. The lower right quadrant depicts evolution with the dynamical spherical AMR criterion, and $N_M=96$ fixed.
    Each grey box (\MeshBlock{}) is sampled with $N_B=16$ points along each axis.
	For the cases involving the pBIB AMR criterion observe that reflection of outward propagating dynamical spherical features in $\tilde{\gamma}_{xx}$ is present in the vicinity of a change in refinement level. See text for further discussion. 
    }
		\label{fig:mesh_calib_110}
\end{figure}

\begin{figure}[!ht]
	\centering
		\includegraphics[width=\columnwidth]{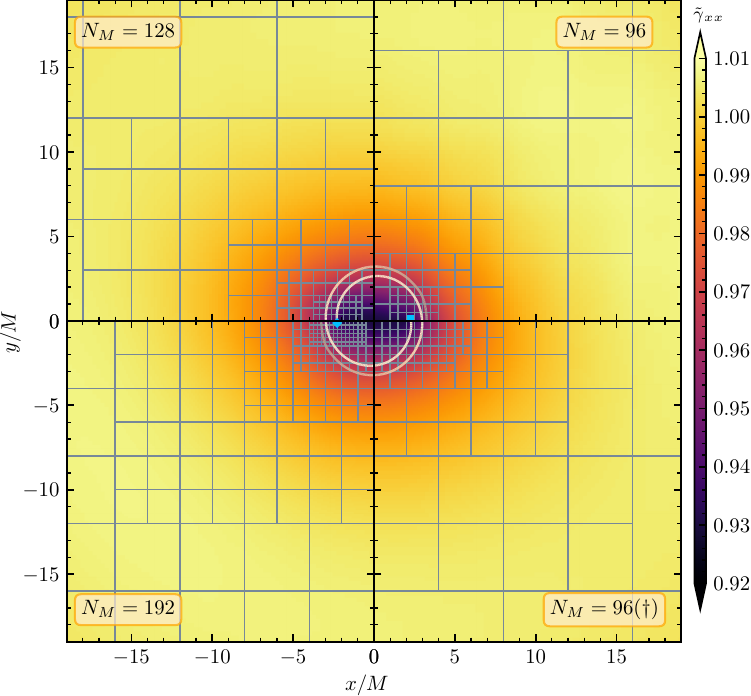}
    \caption{%
	    Zoom of the strong-field region of Fig.\ref{fig:mesh_calib_110}. Trackers are overlaid where lighter color indicates increasing $t$. Observe that the finest resolution in both $N_M=96$ runs is the same, however the region on the finest level is larger in the bottom right (dynamical spherical region) run. Note: color scalings are distinct between this figure and Fig.\ref{fig:mesh_calib_110}.
    }
		\label{fig:mesh_calib_110_zoomed}
\end{figure}

As described in \S\ref{sbsec:meth_ref} AMR conditions may be freely selected. In the approach to a puncture location, the conformal factor $\chi$ approaches a local minimum (see e.g.~\cite{Thierfelder:2010dv}). Indeed, we have numerically verified that the trackers shown in Fig.\ref{fig:tracker_cx_vc} can also be constructed based on such local minimum. On account of the underlying dynamical (approximate) symmetries of the problem, together with their impact on conserved quantities \cite{Mewes:2020vic}, and GW extraction taking place on spheres, an approximately matched AMR condition is motivated. To this end, spherically matched \Mesh{} refinement is induced by fixing levels based on the collection $\{\mathbb{S}{}_{11,3}(\chi^*_\pm),\,\mathbb{S}{}_{6,100}(\mathbf{0})\}$. That is, two spherical regions of radius $3\,M$, with $N_L=11$ follow both punctures. A central fixed sphere of radius $100\,M$, and $N_L=6$ is set to improve resolution in the wave-zone. This is compared against pBIB in Fig.\ref{fig:mesh_calib_110} and Fig.\ref{fig:mesh_calib_110_zoomed}.

\begin{figure}[!ht]
	\centering
		\includegraphics[width=\columnwidth]{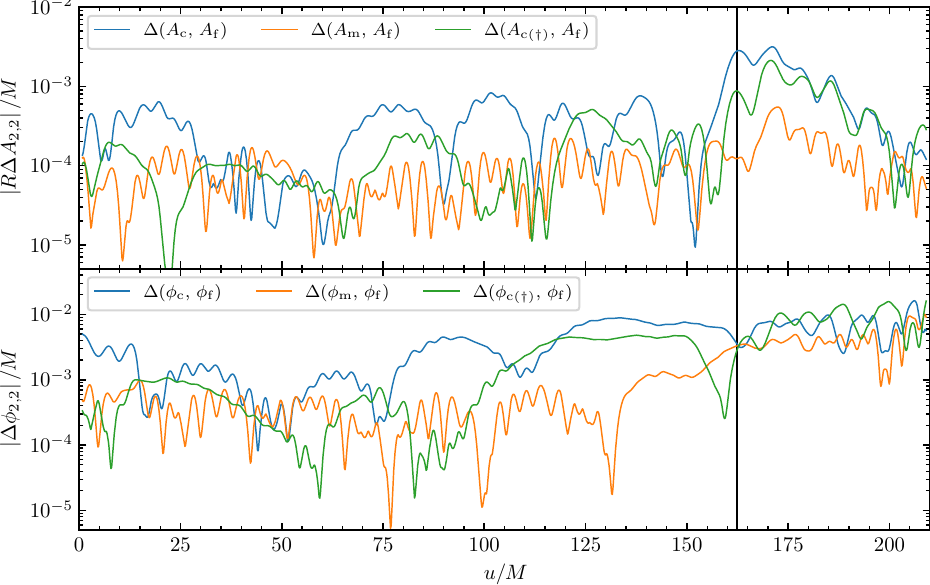}
    \caption{%
      Comparison of differences in the amplitude $A$ of $(2,\,2)$ mode of the gravitational strain (top-panel) and phase $\phi$ (bottom-panel), for runs at different resolutions and refinement strategy, under $\mathcal{Z}_{\mathrm{CX}}$. We consider pBIB AMR with $(c,\,m,\,f)$ corresponding to $N_M\in\{96,\,160,\,192\}$. A single run with spherically matched AMR at $N_M=96$ is denoted with $c(\dagger)$. Vertical black line across panels indicates $u_{\mathrm{merger}}$.
      Extraction performed at $R=100M$.
    }
		\label{fig:wvf_conv_cx_h_2_2_sph}
\end{figure}

Let us consider the behaviour of $\tilde{\gamma}_{xx}$ in the vicinity of a change in refinement level as shown in Fig.\ref{fig:mesh_calib_110}. It is observed that small, spurious reflection occurs, from the underlying rectilinear grid structure. In contrast, by comparing the two $N_M=96$ runs, we see that approximate spherically matched AMR partially mitigates these spurious reflections. The magnitude of the effect can also be seen to converge away when comparing $N_M=192$. The change in refinement strategy also has an effect on GW phasing, which we show in Fig.\ref{fig:wvf_conv_cx_h_2_2_sph}. At early times the amplitude and phasing error of the spherically matched AMR with $N_M=96$ is comparable to that of a pBIB run at $N_M=160$. This can be attributed to a combination of increased wave-zone resolution and better symmetry matching of the AMR (See also the $L_2$ versus $L_\infty$ AMR discussion of \cite{Rashti:2023wfe}).

Another potential advantage of running the spherically matched AMR is a reduction in execution time. Running the spherically matched AMR at $N_M=96$ in bitant results in $\#\mathrm{MB}=4532$ initially, where $\sp_p$ is as in Tab.\ref{tab:CalibBBHgridsLinf}. Compared to pBIB AMR under bitant symmetry reduction at $N_M=160$, it results in a computational resource reduction of $\sim3.4\times$ when coarser time-step $\delta t$ at fixed CFL is taken into account.

Finally, we provide a rough characterization of performance differences between $\mathcal{Z}_{\mathrm{CX}}$ and $\mathcal{Z}_{\mathrm{VC}}$ through a short timing test. To this aim we evolve the problem for the calibration BBH on the pBIB grid configurations of Tab.\ref{tab:CalibBBHgridsLinf}. The initial checkpoint is taken at a time of $t_0=5\,M$, where the AMR condition has refined the \Mesh{}. Timings are as summarized in Tab.\ref{tab:TimingBBH}.
\begin{table}[ht!]
  \centering
  \setlength\tabcolsep{2pt}
  \begin{tabular}{| l || r | r || r ||r | r |}
    \hline
    $N_M$  & $t_F(\mbox{{\tt cx}})$  & $t_F(\mbox{{\tt vc}})$ & $r_t$ ($2\,\mathrm{dp}$) & $\#\mathrm{MB}$ & $\#\mathrm{MB}/\mathrm{Core}$\\
    \hline
    $96$      & $26.48$  & $35.17$    & $0.71$ & $2404$  & $\sim3$\\
    $128$     & $18.01$  & $21.98$    & $0.77$ & $3280$  & $\sim4$\\
    $160$     & $10.41$  & $10.80$    & $0.93$ & $8774$  & $\sim11$\\
    $192$     & $8.49$   & $9.03$     & $0.87$ & $11490$ & $\sim15$\\
    \hline
  \end{tabular}
  \caption{%
    One hour timing test on {\tt SuperMUC-NG} of {\tt LRZ} utilizing $16$ nodes.     The timing ratio $r$ is defined as $r_t:=(t_F(\mbox{{\tt cx}})-t_0)/(t_F(\mbox{{\tt vc}})-t_0)$. The number of {\tt MeshBlock} objects is provided at $t_0$. Bitant symmetry reduction has been performed. The AMR is based on the pBIB criterion.
  }
  \label{tab:TimingBBH}
\end{table}
We see that for pure vacuum, the space-time solver has a distinct advantage when run with the choice $\mathcal{Z}_{\mathrm{VC}}$ in terms of speed. Indeed, we observe an approximate average $\sim20\%$ %
later final evolution time, across $N_M$, when compared to the $\mathcal{Z}_{\mathrm{CX}}$ choice. This we attribute principally to a difference in complexity of the $\mathcal{R}/\mathcal{P}$ operations. A cautionary remark is however in order: the \MeshBlock{}/CPU saturation has not been taken into account, which requires a certain threshold for efficiency \cite{Daszuta:2021ecf,Cook:2023bag}. With regard to performance, these timing results nonetheless strongly suggest that $\mathcal{Z}_{\mathrm {VC}}$ is the better option for BBH.

\subsection{Binary Neutron Stars}
\label{sbsec:mat_tests}
In the case of binary neutron star (BNS) evolution, in addition to the space-time evolution, we need to address the evolution of the hydrodynamical variables. These variables satisfy a balance law of the form $\partial_t[\mathcal{Q}_A] + \partial_i\mathcal{F}{}_A^i[\mathcal{Q}] = \mathcal{S}_A[\mathcal{Q}]$ \cite{Banyuls:1997zz}.
The source term $\mathcal{S}_A[\mathcal{Q}]$ does not involve derivatives of $\mathcal{Q}_A$, but does contain geometric fields and their derivatives. Here the finite-volume, HRSC methods we employ lead us to take the evolved $\mathcal{Q}_A$ on cell-centers. In the $\partial_i\mathcal{F}{}_A^i[\mathcal{Q}] $ approximation step, primitive variables $\mathcal{R}_A[Q]$ are assembled and reconstructed to FC \cite{zanna2002efficientshockcapturingcentraltype}. This process requires geometric data on FC. Information about the evolved fluid variable state, must be available on the grid used for the space-time solver. If we select $\mathcal{Z}_{\mathrm{VC}}$ the required intergrid transfer operations can be summarized as: $\mathcal{Z}_{\mathrm{VC}\rightarrow\mathrm{CC}}$, $\mathcal{Z}_{\mathrm{VC}\rightarrow\mathrm{FC}}$, and finally $\mathcal{R}_{A,\mathrm{CC}\rightarrow\mathrm{VC}}$. On the other hand, selecting $\mathcal{Z}_{\mathrm{CX}}$ means we only need $\mathcal{Z}_{\mathrm{CX}\rightarrow\mathrm{FC}}$.

The intergrid transfer operations summarized above are handled through the use of symmetric Lagrange interpolation. Specifically, during evaluation, $N_I$ nearest neighbor nodes either side of a target interpolation point are used along a salient axis. We have found it beneficial, for stability reasons, to fix $N_I=1$ in a subset of operations. This we refer to as a hybrid mode. In this mode $N_I$ is only allowed to vary freely for $\mathcal{Z}$ derivative assembly entering $\mathcal{S}_A[\mathcal{Q}]$, and primitives $\mathcal{R}_{A,\mathrm{CC}\rightarrow\mathrm{VC}}$ that recouple to the space-time solver.

As initial consistency checks for geometry-matter field coupling and the above operations we consider static star tests in Appendix~\ref{sec:appendix}. A more representative setting of binary neutron star (BNS) evolution problems is required. For this, we consider a quasi-circular, equal-mass BNS inspiral-merger problem. Here irrotational, constraint-satisfying initial data %
is provided by \texttt{Lorene} \cite{Gourgoulhon:2000nn}. Specifically, as a validation problem, we consider the \texttt{G2\_I14vs14\_D4R33\_45km} binary\footnote{%
Dataset available at \url{https://lorene.obspm.fr/}. %
} which features baryon mass $M_b=3.2500\Mo$ and gravitational mass $M=3.0297\Mo$ at an initial separation of $45$~km. The Arnowitt-Deser-Misner (ADM) mass of the binary is $M_{\rm ADM}=2.9984\Mo$, the angular momentum
$J_{\rm ADM}=8.8354\Mo^2$G/c, and the initial orbital frequency is
$f_0\simeq294$~Hz. For these tests we utilize an ideal gas equation-of-state (see Appendix~\ref{sec:appendix}) with $K=123.64$. The evolution of this BNS system proceeds for three orbits prior to the formation of a massive remnant, thereafter leading to gravitational collapse.

The computational domain is again selected as $\Omega:=[-D,\,D]^3$, where $D=1536\,\Msun{}\simeq 2268\,\mathrm{km}$. The AMR is induced through tracking the local minimum of the lapse $\alpha$, as positioned on each constituent of the BNS system. Resolution levels are fixed according to $\{\mathbb{S}_{8,15}(\alpha^*_\pm),\, \mathbb{S}_{7,30}(\alpha^*_\pm),\,\mathbb{S}_{6,110}(\mathbf{0})\}$. To perform a resolution study, we take a sequence of $N_M\in\{64,\,96,\,128\}$. This translates into a sequence of finest resolution levels of $(554,\,369,\,277)\,\mathrm{m}$. The \MeshBlock{} sampling is selected as $N_B=16$.
For the time-evolution we make use of the SSPRK$(3,3)$ method of \cite{gottlieb2009highorderstrong}, with a CFL of $\mathcal{C}_1=0.2$, and $\sigma_{\mathrm{D}}=0.5$. A tenuous, artificial atmosphere is added, where $\rho_{\mathrm{atm}}=10^{-16}$. We do not make use of thresholding \cite{Cook:2023bag}. For the {\tt RePrimAnd}-based primitive recovery we set the tolerance to $10^{-10}$. In this section no symmetry reductions are performed.

\begin{figure}[!ht]
	\centering
		\includegraphics[width=\columnwidth]{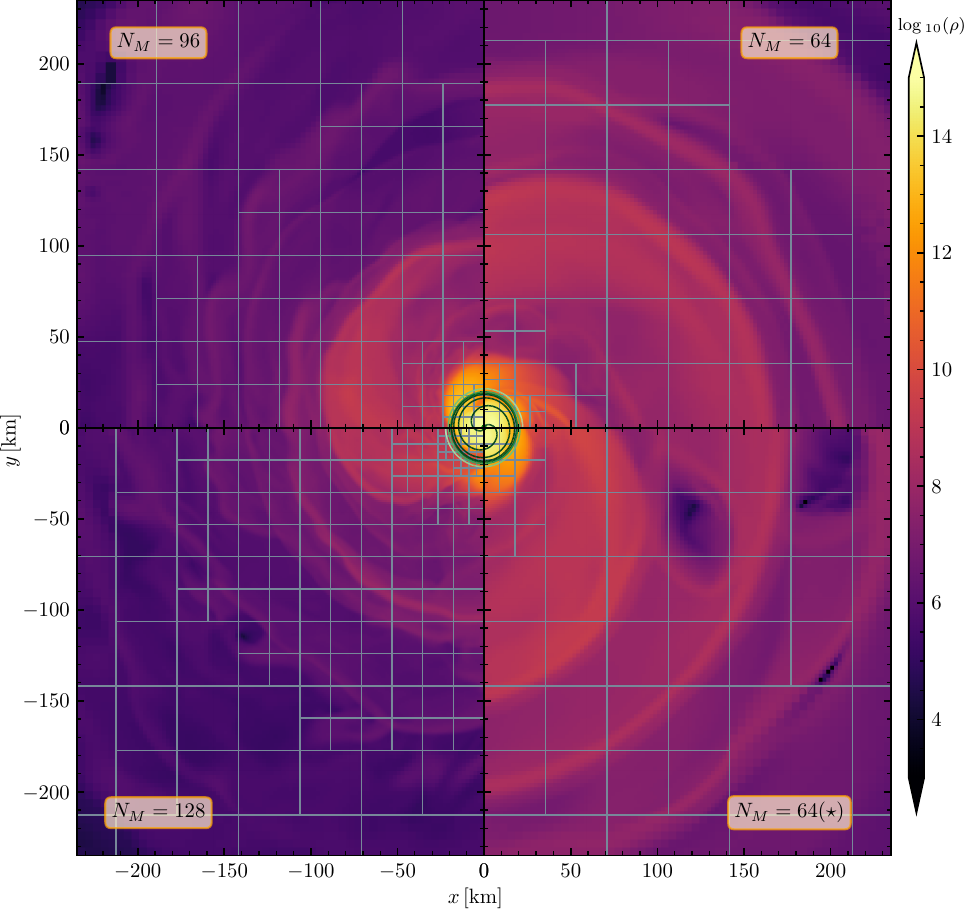}
    \caption{%
    \Mesh{} structure and values of fluid rest mass-density $\rho\,[\mathrm{g}\,\mathrm{cm}^{-3}]$ shortly post-merger
            in the orbital plane for the BNS inspiral problem described in the text. Counter-clockwise, starting from the upper right quadrant we have evolution with CX sampling with $N_M\in\{64,\,96,\,128\}$. The lower right quadrant depicts evolution under VC sampling for geometric fields with $N_M=64$. Each grey box (\MeshBlock{}) is sampled with $N_B=16$. The AMR criterion for each run has driven the distribution of refined region resolution towards an origin-centered approximate sphere. Trackers that follow local minima of $\alpha(t)$ are overlaid. Darker tracker color indicates increasing $t$.
    }
		\label{fig:bns_postmerger_quad}
\end{figure}

We select $\mathcal{Z}_{\mathrm{CX}}$ and evolve the initial data with the aforementioned parameters. An additional run featuring $\mathcal{Z}_{\mathrm{VC}}$ with $N_M=64$ and all else fixed is also performed for internal comparison. In each case we find approximately three orbits prior to the formation of a massive remnant. A snapshot of the fluid rest mass-density at this point ($t\simeq 8.8 \mathrm{ms}$) is shown in Fig.\ref{fig:bns_postmerger_quad} comparing the four runs.
We observe that the spiral arm structure is qualitatively similar for the two runs at $N_M=64$, which provides an initial consistency check. The structure becomes sharper and more compact at higher $N_M$. The extrema tracking based AMR can also be seen to guide increased resolution towards regions that suitably follow the binary consituents.

\begin{figure}[!ht]
	\centering
		\includegraphics[width=\columnwidth]{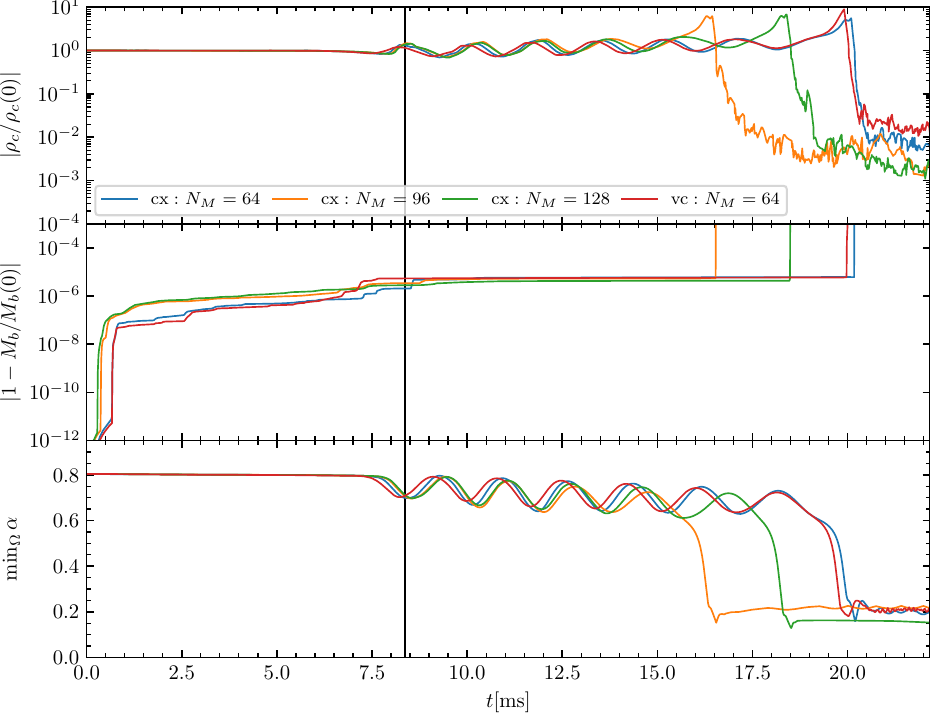}
    \caption{%
    	    Relative change of (global) maximum density $\rho_c$ (upper panel), baryon mass conservation $M_b$ (middle), and minimum of the lapse $\alpha$ (lower panel), for the BNS model problem. A resolution triplet induced by the choices $N_M\in\{64,\,96,\,128\}$ and selection of $\mathcal{Z}_{\mathrm{CX}}$ is depicted, with a single run at $N_M=64$ and $\mathcal{Z}_{\mathrm{VC}}$ shown for reference. Collapse times, for each run, are signalled by: post-merger, global peak in the $\rho_c$ monitor, blow up of the mass conservation monitor, and confirmed by reduction of lapse below $\sim 0.2$.
    Vertical black line depicts merger time ($t\simeq 8.4\,\mathrm{ms}$), as based on the peak amplitude $h{}_{22}$ associated with the $N_M=128$ run (C.f. Fig.\ref{fig:wvf_bns_conv_cx_vc_h_2_2}). See text for further discussion.
    Note: legend common to all panels.
    }
		\label{fig:bns_scalars_I_cmp}
\end{figure}
Quantitative scalar monitors are shown in Fig.\ref{fig:bns_scalars_I_cmp}. We find that the time of collapse falls in the range $t\in[16.5,\,20.2]\,\mathrm{ms}$ for $\mathcal{Z}_{\mathrm{CX}}$ selection. This depends on the selection of $N_M$. Importantly, we observe consistency to within $1\%$ when comparing a verification run with $N_M=64$ based on $\mathcal{Z}_{\mathrm{VC}}$. Violation in baryon mass conservation grows to a plateau of $\sim\mathcal{O}(10^{-6})$ at merger for each run. This remains approximately constant until the point of gravitational collapse. Again we see that both choices of grid sampling for the geometric fields lead to consistent results.

\begin{figure}[!ht]
	\centering
		\includegraphics[width=\columnwidth]{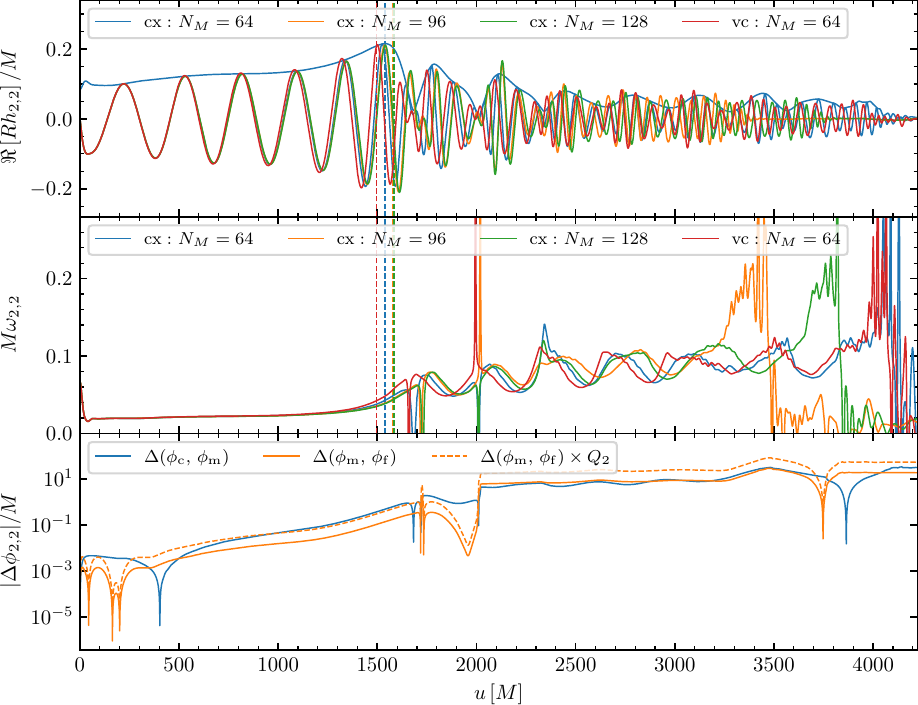}
    \caption{%
    Gravitational strain $(2,\,2)$ mode extracted at $R=100M$ for binary neutron star run of \S\ref{sbsec:mat_tests}. (Top): real part of the strain with $\mathcal{Z}_\mathrm{CX}$ for $N_M\in\{64,\,96,\,128\}$ is depicted. For $N_M=64$ the amplitude is also shown. In the case of $N_M=128$ we find $u_\mathrm{merger}\simeq 1583.2\,M$ ($t_\mathrm{merger}\simeq 1699.8\,M$). We observe consistent behaviour at $N_M=64$ for a comparable run based on VC sampling. (Middle): Instantaneous frequency is shown. (Top, middle): vertical dashed lines indicate time of merger for the different data sets. (Bottom): Phase differences for the CX resolution triplet suggest convergence (to merger) as compatible with a second order trend.
    }
		\label{fig:wvf_bns_conv_cx_vc_h_2_2}
\end{figure}

Gravitational waveforms are the next item to verify. To construct the $(2,\,2)$ mode of the strain, for the BNS system, we use fixed frequency integration (FFI) \cite{Reisswig:2010di}. The FFI proceeds by transforming the signal to the frequency representation, applying a low-frequency cutoff $f_0$, and integrating. We select $f_0=0.002 M^{-1}$. The result is shown in Fig.\ref{fig:wvf_bns_conv_cx_vc_h_2_2}.
Based on the $N_M=128$ run, we find that the time of merger (retarded coordinate) is $u_\mathrm{merger}\simeq 1583.2\,M$. This translates into $t_\mathrm{merger}\simeq 1699.8 \,M\simeq 7.9\,\mathrm{ms}$. Indeed, all CX runs fall within a band of $\sim 0.2\mathrm{ms}$. This indicates strong robustness of the simulations performed. The merger time for the $\mathcal{Z}_{\mathrm{VC}}$ waveform is within $3\%$ that of the CX at $N_M=96$. Similarly, the morphology of the instantaneous frequency for CX when cross-compared against VC is seen to be alike. With the method described in Appendix~\ref{sec:appendix} we also assess convergence. As can be seen in Fig.\ref{fig:wvf_bns_conv_cx_vc_h_2_2} we have a strong indication for a second order trend in the convergence of the $h{}_{22}$ associated phase, as $N_M$ is increased.

Overall, based on this discussion, when running with $\mathcal{Z}_{\mathrm{CX}}$, good agreement is found with $\mathcal{Z}_{\mathrm{VC}}$ here, and our prior results in \cite{Cook:2023bag}. This provides confidence in the correctness of our implementation, and the validity of the strategy outlined in \S\ref{sbsec:meth_ref}.

To close this section, as in the BBH case, we provide a rough characterization of performance differences between $\mathcal{Z}_{\mathrm{CX}}$ and $\mathcal{Z}_{\mathrm{VC}}$ through a short timing test. The initial checkpoint is taken at a time of $t_0=5\,M$ with timings summarized in Tab.\ref{tab:TimingBNS}.

\begin{table}[ht!]
  \centering
  \setlength\tabcolsep{2pt}
  \begin{tabular}{| l || r | r || r ||r | r |}
    \hline
    $N_M$  & $t_F(\mbox{{\tt cx}})$  & $t_F(\mbox{{\tt vc}})$ & $r_t$ ($2\,\mathrm{dp}$) & $\#\mathrm{MB}$ & $\#\mathrm{MB}/\mathrm{Core}$\\
    \hline
    $64$      & $170.45$ & $151.02$   & $1.13$ & $3676$  & $\sim 5$\\
    $96$      & $70.84$  & $61.11$    & $1.17$ & $6768$  & $\sim 9$\\
    $128$     & $34.78$  & $28.36$    & $1.27$ & $12384$ & $\sim 16$\\
    \hline
  \end{tabular}
  \caption{%
    One hour timing test on {\tt SuperMUC-NG} of {\tt LRZ} utilizing $16$ nodes.     The timing ratio $r$ is defined as $r_t:=(t_F(\mbox{{\tt cx}})-t_0)/(t_F(\mbox{{\tt vc}})-t_0)$. The number of {\tt MeshBlock} objects is provided at $t_0$. No symmetry reduction has been performed. The AMR is based on spherical regions criteria.
  }
  \label{tab:TimingBNS}
\end{table}
We see that for BNS, the space-time solver has a distinct advantage when run with the choice $\mathcal{Z}_{\mathrm{CX}}$ in terms of speed. Indeed, we observe an approximate average $\sim17\%$ later evolution time, across $N_M$, when compared to the $\mathcal{Z}_{\mathrm{VC}}$ choice. This is in direct contrast to the BBH case (Tab.\ref{tab:TimingBBH}). While it is the case that the level-to-level operators become more involved for CX, when contrasted against VC, the reduction in the number of intergrid transfer operations appears to compensate. Indeed, it appears that the gap between CX and VC ($r_t$ of Tab.\ref{tab:TimingBNS}) increases with increasing $N_M$. This is also in contrast to the BBH case, where the gap appeared to decrease. Overall, we see that CX achieves later final evolved times in the BNS problem. Hence we conclude that $\mathcal{Z}_{\mathrm{CX}}$ is the favoured choice, for this class of problem.

\section{Conclusion}
\label{sec:discussion}
In this work we have proposed a new, cell-centered (CX) variable treatment within the context of our space-time solver embedded within \GRAthena{}. This has allowed us to solve the \z4c{} system in $\mathcal{Z}_{\mathrm{CX}}$ mode, in a fashion consistent with our earlier work on vertex-centers, i.e.~$\mathcal{Z}_{\mathrm{VC}}$ mode. Tests for both vacuum and GRHD problems demonstrated robust functionality for either grid sampling. The choice of $\mathcal{Z}_{\mathrm{CX}}$ for GR(M)HD yields a substantial simplification in treatment of discretized variables.

To validate the new treatment, together with implementation details within the code, first consistency tests based on BBH evolution were performed. A step-by-step strategy was taken which featured, inter alia, a resolution study. This allowed us to assess a trend compatible with $\6th$ order convergence in the error of the phase of the $(2,\,2)$ mode of the gravitational wave strain, when working with evolution of $\mathcal{Z}_{\mathrm{CX}}$. This is consistent with our prior work \cite{Daszuta:2021ecf}. In order to provide insight as to which grid sampling strategy performs better, for BBH evolution, we considered a one hour timing test on {\tt SuperMUC-NG} of the Leibniz-Rechenzentrum Munich. It was found that $\mathcal{Z}_{\mathrm{VC}}$, on average, is $\sim20\%$ faster. This we attribute to the differences in how the level-to-level refinement operators are evaluated for the two grid samplings. Motivated by the ``symmetry-seeking'' property of the $L_2$-AMR previously investigated in \cite{Rashti:2023wfe}, we proposed, a simplified, spherical-region based AMR strategy. This was demonstrated in the BBH validation problem to potentially partially mitigate some rectilinear grid imprinting.

Following this, a quasi-circular, equal-mass, BNS inspiral, merger, and gravitational collapse problem was illustrated in the context of GRHD. We again utilized the $(2,\,2)$ mode of the gravitational wave strain as a consistency diagnostic. A trend compatible with $\2nd$ order convergence was assessed in a resolution study when working with $\mathcal{Z}_{\mathrm{CX}}$. This trend is consistent with that observed in \cite{Cook:2023bag} when working with $\mathcal{Z}_{\mathrm{VC}}$. With regard to performance, in contrast to the BBH case, it was found that $\mathcal{Z}_{\mathrm{CX}}$ is, on average, $\sim17\%$ faster for the runs considered. This we attribute to a reduction in the total number of intergrid interpolation operations required. For these tests, we also leveraged the simplified, spherical-region based AMR strategy. While we do not explicitly demonstrate GRMHD runs in this work, we have found another potential advantage of working with $\mathcal{Z}_{\mathrm{CX}}$. First tests, involving the initial data of the BNS simulation presented here, allowed for evolution with magnetic fields through merger and collapse, \textit{without} apparent horizon-based excision. This contrasts against the situation with $\mathcal{Z}_{\mathrm{VC}}$ where we have found such excision techniques to be crucial.

\begin{acknowledgments}
  BD and SB acknowledges support by the EU Horizon under ERC Consolidator Grant,
  no. InspiReM-101043372. 
  PH acknowledges funding from the National Science Foundation under Grant No.~PHY-2116686.
  EG, DR acknowledge funding from the National Science Foundation under grant No.~AST-2108467.
  EG acknowledges funding from an Institute for Gravitation and Cosmology fellowship.
  JF, DR acknowledge U.S. Department of Energy, Office of Science, Division of
  Nuclear Physics under Award Number(s) DE-SC0021177.
  DR acknowledge support from NASA under award No.~80NSSC21K1720 and
  from the Sloan Foundation.
  The authors are indebted to A.~Celentano's PRİSENCÓLİNENSİNÁİNCIÚSOL.

  The numerical simulations were performed on the national HPE Apollo Hawk 
  at the High Performance Computing Center Stuttgart (HLRS). 
  The authors acknowledge HLRS for funding this project by providing access 
  to the supercomputer HPE Apollo Hawk under the grant numbers INTRHYGUE/44215
  and MAGNETIST/44288. 
  Simulations were also performed on SuperMUC\_NG at the
  Leibniz-Rechenzentrum (LRZ) Munich.  
  The authors acknowledge the Gauss Centre for Supercomputing
  e.V. (\url{www.gauss-centre.eu}) for funding this project by providing
  computing time on the GCS Supercomputer SuperMUC-NG at LRZ
  (allocations {\tt pn68wi} and {\tt pn36jo}).
  Simulations were also performed on TACC's Frontera (NSF LRAC
  allocation PHY23001) and on Perlmutter. This research used resources
  of the National Energy Research Scientific Computing Center, a DOE
  Office of Science User Facility supported by the Office of Science of
  the U.S.~Department of Energy under Contract No.~DE-AC02-05CH11231.
  Postprocessing and development run were performed on the ARA cluster
  at Friedrich Schiller University Jena. 
  The ARA cluster is funded in part by DFG grants INST
  275/334-1 FUGG and INST 275/363-1 FUGG, and ERC Starting Grant, grant
  agreement no. BinGraSp-714626.
\end{acknowledgments}

\onecolumngrid
\appendix
\section{Assessing convergence with a static star}
\label{sec:appendix}

\begin{figure}[!ht]
	\centering
		\includegraphics[width=0.5\columnwidth]{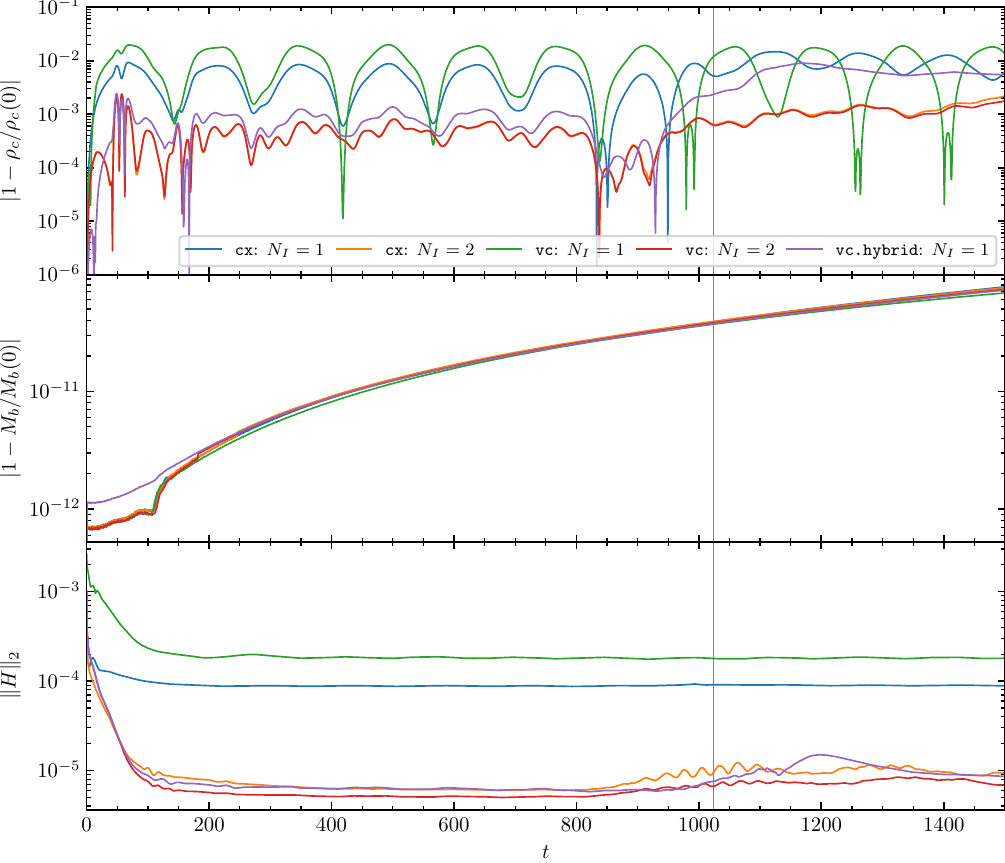}
    \caption{%
      Comparison of scalar quantities in stable star evolution (TOV test) on a fixed grid configuration involving CX and VC sampling where various varieties of intergrid interpolation of \S\ref{sbsec:mat_tests} are utilized. The \Mesh{} and \MeshBlock{} samplings are $N_M=64$ and $N_B=16$ respectively. The computational domain is taken as $\Omega:=[-1024,\,1024]^3$ with a central, cubic, region $\Omega_r:=[-10,\,10]^3$ refined with $N_L=7$. The region $\Omega_r$ completely contains the star. As intergrid interpolation order is increased it can be seen (upper panel) that central density is better preserved, with oscillation magnitude of $\rho_c(t)$ reduced. Hamiltonian constraint violation (lower panel) is also reduced as interpolation order is increased. We find common mass conservation (middle panel) properties after $\sim t = 200$ regardless of interpolation order, or geometric sampling. Vertical grey line indicates first time of causal contact of $\partial\Omega$ with the origin.
    }
		\label{fig:tov_scalars_I_cmp}
\end{figure}

Evolving a stable, static star in the general relativistic setting provides for a simple, sanity check on code correctness. To this end initial data (ID) is prepared based on the assumption of a spherically symmetric, self-gravitating matter distribution at equilibrium satisfying the Tolman-Oppenheimer-Volkoff (TOV) equations (see e.g.~\cite{Font:1998hf}). For ID preparation a polytropic EOS $p=K \rho^\Gamma$ is assumed with adiabatic index $\Gamma=2$, and polytropic constant $K=100$ as in \cite{Bernuzzi:2009ex}. The stable star model is taken to have a central density $\rho_c=1.280\times 10^{-3}$, gravitational mass $M=1.400$, and circumferential radius $R=9.586$. During evolution we make use of the more general EOS $p=(\Gamma-1)\rho\epsilon$ with $\epsilon=K \rho^{\Gamma-1}/(\Gamma-1)$. 
We perform an evolution of this data to a final time of $T=1500$. This is done for both $\mathcal{Z}_{\mathrm{VC}}$ and $\mathcal{Z}_{\mathrm{CX}}$ choices, for a variety of intergrid interpolation orders (see \S\ref{sbsec:mat_tests}). For these runs we have taken a CFL of $\mathcal{C}_1=0.25$, Kreiss-Oliger dissipation $\sigma_{\mathrm{D}}=0.5$, and set artificial atmosphere to $\rho_{\mathrm{atm}}=10^{-18}$. The \z4c{} damping parameters are taken as $(\kappa_1,\,\kappa_2)=(0.02,\,0)$. Scalar diagnostic quantities are shown in Fig.\ref{fig:tov_scalars_I_cmp}.

To assess solution convergence we consider numerical evolution repeated at a triplet of coarse, medium, and fine resolutions $(\delta x_c,\, \delta x_m,\, \delta x_f)$ that satisfy $\delta x_c > \delta x_m > \delta x_f$. 
The convergence rate of an approximation to the corresponding field data $\mathcal{F}$ may be investigated by comparing differences of solutions $\delta\mathcal{F}_{ab}:=\mathcal{F}_a-\mathcal{F}_b$ at distinct resolutions. For $\delta x^n$ one finds based on Taylor expansion that $\delta\mathcal{F}_{cm} \simeq \delta\mathcal{F}_{mf} \times Q_n(c,m,m,f)$ where we have introduced the so-called convergence factor:
\begin{eqnarray}
  \label{eq:conv_fact}
  Q_n(a,\,b,\,c,\,d) := \frac{\delta x_a^n - \delta x_b^n}{\delta x_c^n - \delta x_d^n}.
\end{eqnarray}

\begin{figure}[!ht]
	\centering
		\includegraphics[width=0.5\columnwidth]{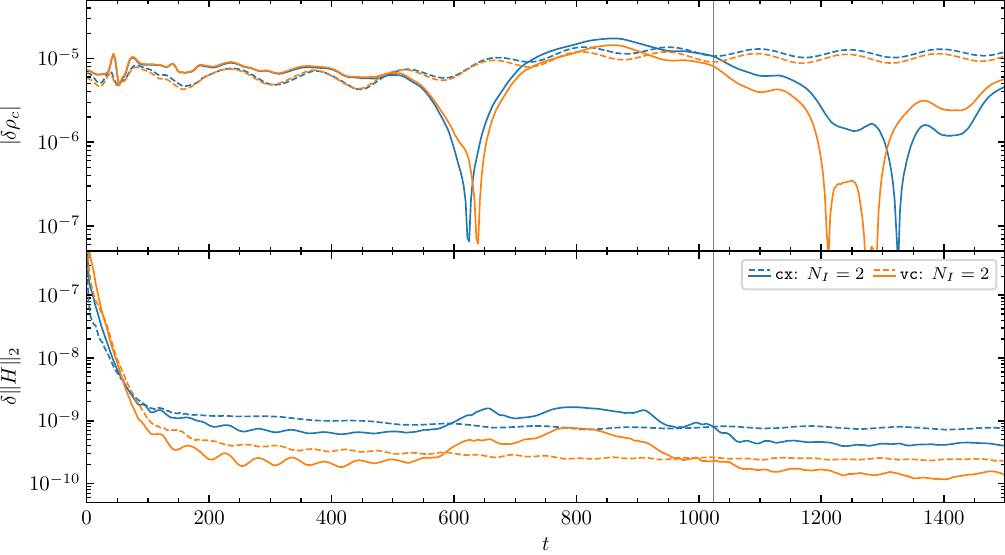}
    \caption{%
      Convergence of scalar quantities in the TOV test for a fixed grid configuration involving CX and VC sampling. Overall grid structure is as in Fig.\ref{fig:tov_scalars_I_cmp} where resolution triplets are constructed based selecting the level of $\Omega_r$ refinement through $N_L\in\{5,\,6,\,7\}$. In dashed lines we depict the rescaled coarse-medium difference according to $Q_n$ of Eq.\eqref{eq:conv_fact}. As additionally verified through least-squares fitting on $t\in[128,\,950]$ we find: $n\sim2$ i.e.~$\sim 2^{\mathrm{nd}}$ order convergence for $\rho_c$ (upper panel); $n\sim5$ for the Hamiltonian constraint (lower panel).
    }
		\label{fig:tov_scalars_I_conv}
\end{figure}

The $N_I=2$ simulations featuring CX and VC of Fig.\ref{fig:tov_scalars_I_cmp} show the smallest deviation from the initial value $\rho_c$ during evolution, together with the smallest constraint violation. We construct a triplet of resolutions for convergence assessment by repeating the runs with $N_L\in\{5,\,6\}$. Based on the refinement structure this entails $(\delta x_c,\, \delta x_m,\, \delta x_f)=(\delta x_c,\, \delta x_c/2,\, \delta x_c/4)$ on the finest level, covering the star, for each run. Behaviour for the central density, together with the Hamiltonian constraint for our convergence assessment is shown in Fig.\ref{fig:tov_scalars_I_conv}. As can be seen we observe ~$\sim 2^{\mathrm{nd}}$ order, and $\sim 5^{\mathrm{th}}$ order compatible trends for the aforementioned quantities respectively.
Based on the consistency between the qualitative features of the CX and VC scalar diagnostics depicted in Fig.\ref{fig:tov_scalars_I_cmp}, together with consistent order of convergence demonstrated in Fig.\ref{fig:tov_scalars_I_conv} we consider the TOV test to pass.
\end{document}